\documentclass[12pt]{article}
\usepackage{pdproc} 
\usepackage{axodraw}
\usepackage{graphicx}
\usepackage{epsfig}

  \newcommand{\be}{\begin{equation}}
   \newcommand{\ee}{\end{equation}}
     \newcommand{\bea}{\begin{eqnarray}}
   \newcommand{\eea}{\end{eqnarray}}

\begin{document}

\title{ Universality of the Weak Interactions, Cabibbo theory and where they led us\footnote {\small Talk delivered at Pennsylvania University, Philadelfia, April 27, 2011, on the occasion of the B. Franklin Prize 2011, attributed to Nicola Cabibbo, {\it for his key contributions to understanding the symmetries underlying the decays of elementary particles by weak interactions}.}\footnote{Published in Rivista del Nuovo Cimento, {\bf 34}, 679 (2011) }} 
\author{ Luciano MAIANI}
% \maketitle
\address{ Dipartimento di Fisica, Universita' di Roma La Sapienza,
  Piazza A. Moro 1\\
 00185 Roma, Italia\\
 {\rm E-mail: luciano.maiani@roma1.infn.it}
 }

\abstract{Universality of the weak interactions is reviewed, with special emphasis on the origin of the Cabibbo theory  of strange particles $\beta-$decays and its role in the discovery of the unified Electroweak Theory. Achievements and present challenges of the Standard Theory of particles interactions are briefly illustrated. As an homage to Nicola Cabibbo,  his leading role in the Roma school of theoretical physics and in the italian science in general is reviewed.  A selection of papers by Cabibbo and other authors, reprinted from {\it Il Nuovo Cimento} and historically related to the arguments considered here, is presented. The picture is completed with the classical paper by Cabibbo and Gatto on electron-positron collisions and Cabibbo's paper on the weak interaction angle, reprinted from {\it Physical Review} and {\it Physical Review Letters}, respectively.}

\normalsize\baselineskip=15pt
 
\section{Universal Weak Interactions}

In a 1961 book, Richard Feynman\cite{feyn1} vividly described his and Murray Gell-Mann's satisfaction at explaining the close equality of the muon and neutron beta decay Fermi constants. 
They\cite{FGM1} and, independently, Gershtein and Zeldovich\cite{GZEL} had discovered the universality of the weak interactions, closely similar to the universality of the electric charge and a tantalising hint of a common origin of the two interactions. 
But Feynman recorded also his disconcert following the discovery that the Fermi constants of the strange particles, e.g. the $\beta$-decay constant of the $\Lambda$ baryon, turned out to be smaller by a factor of 4-5. 
It was up to Nicola Cabibbo\cite{weakangle} to reconcile strange particle decays with the universality of weak interactions, paving the way to modern electroweak unification. 

\section{Nicola Cabibbo: The beginning} 
Cabibbo's scientific life, {\it first steps}:
\begin{itemize}
\item
graduates in 1958, tutor Bruno Touschek;
\item
becomes the first thoretical physicist in Frascati, hired by G. Salvini;
\item
meets there Raoul Gatto (5 years elder) who was coming back from Berkeley and begins an extremely fruitful collaboration;
\item
witnesses exciting times in Frascati: the first $e^+ e^-$ collider, AdA (Anello di Accumulazione = storage ring), to be followed, later, by larger machine, Adone (= larger AdA), reaching up to 3 GeV in the center of mass (= laboratory) frame; new particles (the $\eta$ meson) studied at the electro-synchrotron, related to the newly discovered SU(3) symmetry, etc.;
\item
publishes together with Gatto an important article on $e^+ e^-$ physics\cite{GCelpos} (the Bible);
\item
in 1961, again with Gatto, investigates the weak interactions of hadrons in the framework of the newly discovered SU(3) symmetry.
\end{itemize}

\section{The V-A and Current x Current theory of the Weak Interactions}

The Fermi weak interaction lagrangian was simply the  product of four fermion fields $\psi_i$ connected by Dirac matrices, which Fermi, to keep the analogy with electromagnetism, restricted to be $\gamma_\mu$ matrices. For the neutron $\beta$-decay:
\be
{\cal L}_{n}= G\left[{\bar \psi_p}\gamma_\mu \psi_n \right]\times \left[{\bar \psi_e}\gamma^\mu \psi_\nu \right]\;+\; {\rm h.c.}
\label{Fermi1}
\ee

Subsequent studies of nuclear decays and the discovery of parity violation, led to complicate the gamma matrix structure, introducing all possible kinds of relativistically invariant products of two fermion fields bilinear. At the end of the fifties, simplicity finally emerged, with the recognition that all $\beta$-decays could be described by a V-A theory.
Sudarshan and Marshak\cite{MarSud}, and Feynman and Gell-Mann\cite{ancora} proposed the general rule:
\begin{itemize}
\item
{\it every $\psi$ replaced by $a\psi$, with: $a=\frac{1-\gamma_5} {2}$.}
\end{itemize}
With this position, we are brought essentially back to Fermi. The lagrangian in (\ref {Fermi1}) reads now:
\be
{\cal L}_{n}= \frac{G}{\sqrt{2}}\left[{\bar \psi_p}\gamma_\mu(1-\gamma_5) \psi_n \right]\times \left[{\bar \psi_e}\gamma^\mu(1-\gamma_5) \psi_\nu \right]
\label{Fermi2}
\ee
(the factor $1/\sqrt{2}$ is inserted so as to keep the constant $G$ at the same value determined by Fermi from superallowed nuclear transitions). 

The V-A structure in Eq.(\ref{Fermi2}) is {\it almost} experimentally correct. The coefficient of $\gamma_5$ in the nuclear bilinear is in fact $g_A/g_V \simeq 1.25$ rather than unity, to be interpreted as a strong interaction renormalisation.

Under the $(1-\gamma_5)$ rule given above, only vector and axial vector currents survive in the Fermi interaction. Eq.(\ref{Fermi2}) further suggests the Current $\times$ Current hypothesis:
\begin{itemize}
\item
{\it the lagrangian of the full weak interactions, describing muon, meson etc. $\beta$-decays, has the form:}
\be
{\cal L}_{W}=  \frac{G}{\sqrt{2}}\; J^\mu \times J_\mu^+,
\label{JJ}
\ee
\end{itemize}
%, namely that the lagrangian of the full weak interactions, describing muon, meson etc. $\beta$-decays, has the form:
%\be
%{\cal L}_{W}=  \frac{G}{\sqrt{2}}\; J^\mu \times J_\mu^+
%\label{JJ}
%\ee
with $J_\mu$ the sum of $n-p$, $e-\nu_e$, etc. contributions. Omitting gamma matrices:
\be
J= (\bar \nu_e e) + (\bar \nu_\mu \mu) + (\bar p n) + X.
\label{current}
\ee

$X$ represents the contribution of the current to strange particle decays and we have to consider now what properties the term X might have (I follow here almost {\it verbatim} the considerations made by Feynman in \cite{feyn1}).

A first observation is that if we insert the form (\ref{current}) into (\ref{JJ}), the terms corresponding to electronic and muonic decays of strange particles will appear with the same coefficient. This corresponds to the so-called {\it electron-muon universality}, which indeed is very well satisfied in strange particle $\beta$-decays.

Second, semi-leptonic decays of strange particles seem to be suppressed with respect to nuclear $\beta$-decays, which implies the term $X$ to appear with a small coefficient, of the order of $0.1$. 

However, if that were the case, a similar suppression should hold for the term $X\times (\bar n p)$, which, judging from $K_S$ decay does not seem to be the case.

Here ends Feynman's analysis of 1961. 
In modern terms, the suppression of the semi-leptonic strange particle decays got mixed with the {\it $\Delta I = 1/2$ enhancement} of non-leptonic decays, resulting in what seemed to be, at the time, a really inextricable mess.

\section{Gell-Mann and Levy's ansatz}

An observation made in 1960 by M. Gell-Mann and M. Levy \cite{GML} is often quoted as a precursor or source of inspiration for Cabibbo. This is justified to some extent, but the role of Gell-Mann and LevyÕs observation need not be overestimated. The Gell-Mann and Levy's paper is quoted by Cabibbo and was well known to all those working in the field.

In the GML paper, the weak current is written in the Sakata model, with elementary P, N and $\Lambda$. All hadrons are supposed to be made by these three fundamental fields. 
GML observe that one could relate the reduction of the $\Lambda$ coupling w.r.t. the muon coupling by assuming the following form of the weak vector current:

\be
V_\lambda=\frac{1}{\sqrt{1-\epsilon^2}}\left[{\bar P}\gamma_\lambda \left(N+\epsilon \Lambda \right) \right]
\label{GML}
\ee

But$...$ nobody knew how to proceed from the GML formula to a real calculation of meson and baryon decays, for two reasons:

{\bf i}. The Sakata model was already known to be substantially wrong, due to the absence of positive-strangeness barions. Thus, inclusion of the decays of the S=-1 and S=-2 hyperons was completely out of reach.

{\bf ii}. The important point of the non-renormalisation was missed. In Gell-Mann and Levy's words\cite{GML}: {\it There is, of course, a renormalization factor for that decay,  (i.e. $\Lambda$ decay) so we cannot be sure that the low rate really fits in with such a picture}.

\section{SU(3) Symmetry and weak interactions}

Gatto and Cabibbo\cite{GC61} and Coleman and Glashow\cite{CG61} observed that the Noether currents associated to the newly discovered SU(3) symmetry include a strangeness changing current that could be associated with strangeness changing decays, in addition to the isospin current responsible for strangeness-non-changing  beta decays (CVC). 
The identification, however, implied the rule $\Delta S= \Delta Q$ in the decays, in conflict with some alleged evidence of a $\Delta S=-  \Delta Q$ component, indicated by the single event $\Sigma^+\to \mu^+ + \nu + n$  reported in an emulsion experiment\cite{ABG62}. In addition, the problem remained how to formulate correctly the concept of CVC and muon-hadron universality in the presence of {\it three }Noether currents:
    
\begin{eqnarray}
&&V_\lambda^{lept}={\bar \nu}_\mu\gamma_\lambda \mu+{\bar \nu}_e \gamma_\lambda e~{\rm (\Delta Q=1)}
\label{muone}\\
&&V_\lambda^{(1)}+i V_\lambda^{(2)}~{\rm (\Delta S=0,~\Delta Q=1)}
\label{s=0} \\
&&V_\lambda^{(4)}+i V_\lambda^{(5)}~{\rm (\Delta S=\Delta Q=1)}
\label{s=1}
\end{eqnarray}

\section{Enters Cabibbo}
In his 1963 paper, Nicola made a few decisive steps. 
\begin{itemize}
\item
he decided to ignore the evidence for a $\Delta S=- \Delta Q$ component. Nicola was a good friend of Paolo Franzini, then at Columbia University, and the fact that Paolo had a larger statistics without any such event was crucial; 
\item
he ignored also the problem of the normalisation of non-leptonic processes and of the $\Delta I=1/2$ enhancement; 
\item
he formulated a notion of universality between the leptonic current and one, and only one, hadronic current, a combination of the SU(3) currents with $\Delta S=0$ and $\Delta S=1$: the hadronic current has to be equally normalized to each component of the lepton current (electronic or muonic). Axial currents are inserted via the V-A hypothesis. 
\end{itemize}

 In formulae, Cabibbo wrote:
\be
V_\lambda^{(hadron)}=a\left[V_\lambda^{(1)}+i V_\lambda^{(2)}\right]+b\left[V_\lambda^{(4)}+i V_\lambda^{(5)}\right]
\label{hadron}
\ee
with 
\be
a^2+b^2=1
\label{normaliz}
\ee
to ensure equal normalization of the hadronic with respect to either the electron or the muon component of the leptonic vector current, Eq.(\ref{muone}). 

Adding these hypotheses to the V-A formulation of the weak interactions, Cabibbo thus arrived to the final expression of the total leptonic and hadronic weak currents:
\bea
&&J_\lambda^{lept}={\bar \nu}_\mu\gamma_\lambda (1-\gamma_5)\mu +{\bar \nu}_e\gamma_\lambda (1-\gamma_5)e;
\label{totmuone}\\
&&J_\lambda^{(hadron)}=\cos\theta\left[J_\lambda^{(1)}+i J_\lambda^{(2)}\right]+\sin\theta \left[J_\lambda^{(4)}+i J_\lambda^{(5)}\right];\label{tothadr}\\
&&J_\lambda^{(i)}=V_\lambda^{(i)}-A_\lambda^{(i)}\label{VmenoA}
\eea

In the above equations, $A_\lambda^{(i)}$ denotes an octet of axial-vector currents. While the normalization of the vector currents is fixed by the very notion of CVC, the axial currents are not conserved and their normalization constants are free parameters, not determined by the $SU(3)$ symmetry%\cite{commento}
. The angle $\theta$ is a new constant of Nature, since known as the {\it Cabibbo angle}.

In the Cabibbo theory:
\begin{itemize}
\item
Currents belong to SU(3)$\times$SU(3);
\item
Partial conservation of the vector and axial vector currents protects the normalization of strenght;
\item
the Gatto-Ademollo  theorem\cite{GAd} holds: vector current matrix elements are not renormalized to first order in SU(3) breaking.
\end{itemize}

The phenomenological success of the Cabibbo theory for semileptonic decays has made it clear that the I=1/2 enhancement of non-leptonic decays must have a different origin than the normalization of the strange particle current, X. This was understood later as a renormalization group effect, as first guessed by K. Wilson\cite{kwils} and computed in QCD by M. K. Gaillard and B. W. Lee and by G. Altarelli and L. Maiani\cite{nonlept}.

As of today, the  agreement of the Cabibbo theory with experiments has been but reinforced by the most recent data from Frascati, FermiLab and CERN\cite{PDG}.

\section{The weak current of baryons and the unitarity limit}

The form of $J_\lambda^{(hadron)}$, well readable in terms of the SU(3) symmetry, leads to a remarkably complicated form of the current in terms of individual baryon fields (to be compared with the Gell-Mann and Levy's form).:
\begin{eqnarray}
&& J_\mu^{(had)}=\cos\theta ~\bar{p}\gamma_\mu\left[1 -(F+D)\gamma_5\right] n +\sin\theta \left\{-\sqrt{\frac{3}{2}}\bar{p}\gamma_\mu\left[1 -(F+\frac{1}{3}D)\gamma_5\right]\Lambda\right\}+ \nonumber
\\
&&+\sin\theta\left\{-\bar{n}\gamma_\mu\left[1 -(F-D)\gamma_5\right]\Sigma^- -\bar{\Sigma^+}\gamma_\mu\left[1 -(F+D)\gamma_5\right]\Xi^0\right\}+ \nonumber \\
&& +\sin\theta\left\{ \sqrt{\frac{3}{2}}\bar{\Lambda}\gamma_\mu \left[1 -(F-\frac{1}{3}D)\gamma_5\right]\Xi^- \right\}+ \nonumber \\
&& +\cdots %\nonumber
\label{cabibbobaryons}
\end{eqnarray}
We have used particle's names to indicate the corresponding fields; F and D are phenomenological coefficients related to axial current renormalization, see the comment made after Eq.(\ref{Fermi2}). 
Experimental data require\cite{cabibboetal}: $F\simeq 0.46$; $D\simeq 0.80$; sin$\theta \simeq 0.22$.

The first term in Eq.(\ref{cabibbobaryons}) describes the $\Delta S=0$, $n\to p$ transition and is normalized by the factor cos$\theta$ which is, of course, less than unity. Thus the Cabibbo theory may explain the observed reduction of the nuclear Fermi constant with respect to the muon one, a fact noticed already by Feynman in \cite{feyn1} following the precise measurement by V. Telegdi and coworkers\cite{Burgy:1960zz}. The effect was not so clear at that time, as it had to be disentangled by competing electromagnetic radiative corrections, which were not under control in the early sixties. The situation is much more clear today, with precise data coming from superallowed Fermi nuclear transition and radiative corrections under control.  

\begin{figure}[htb]
\begin{center}
\includegraphics[scale=.45]{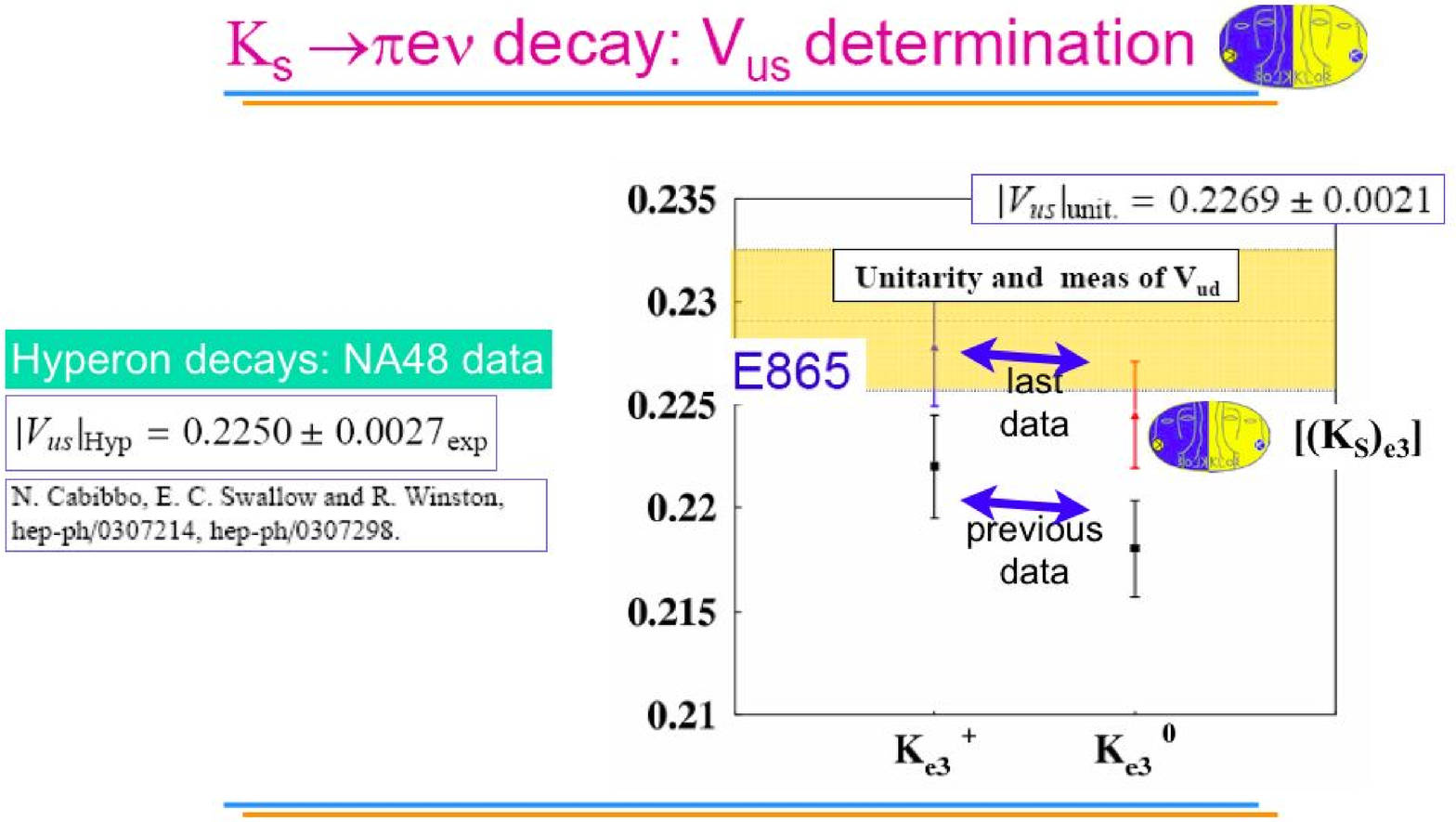}
\caption{
{\small 
Test of Cabibbo unitarity. The yellow band indicates the range of $V_{us}$=sin$\theta$ predicted from the value of $V_{ud}$=cos$\theta$ measured in superallowed nuclear transitions. Indicated are also the latest values of sin$\theta$ obtained from $Kl3$ decays by the experiments E865 (Fermilab) and KLOE (Frascati); the value of $V_{us}$ from strange hyperon decays%\cite{cabibboetal} 
is also reported.}
}
\label{univ}
\end{center}
\end{figure}

As shown in Fig. \ref{univ}, the determination of the angle from the baryonic $\Delta S=1$ and the latest data on $Kl3$ decays presented by the Fermilab, E865, and Frascati, KLOE, experiments, agree {\it extremely} well with what predicted from the superallowed nuclear transitions\cite{Miscetti}.

%
%%%%%%%%%%%%@@@
%%%%%%%%%@@@@
%\vskip0.5cm%

\section{Cabibbo theory with quarks}

Gell-Mann-Levy's formula was given a new life in the context of the quark model, after the consolidation of the Cabibbo theory. If quarks and flavor-singlet gluons are the fundamental particles, as we know today, $\beta$-decays of baryons and mesons simply reflect the two transitions:
 \be
 d\to u, ;\ s\to u
 \ee 
 Note that this is similar to Fermi's idea that $\beta$-decays of nuclei are simply the manifestation of the $n\to p$  transition.
 
 In the quark picture, the Cabibbo weak current takes the form:
\begin{eqnarray}
&&J_\lambda=\cos\theta\left[{\bar u}\gamma_\lambda(1-\gamma_5)\left(d+\tan\theta s \right)\right]=\nonumber \\
&&={\bar u}\gamma_\lambda(1-\gamma_5)d_C
\label{Cabibbo2}
\end{eqnarray}
which coincides with Gell-Mann and Levy's with: $(P, N, \Lambda)\to(u, d, s)$.  
The Cabibbo angle, $\theta$, is seen as the mixing angle expressing the weakly interacting down-quark, $d_C$, in terms of the mass-eigenstate fields: $d, s$.

\section{Equal normalization ?}

It was clarified by Cabibbo himself, in his 1964 Erice lectures, that the condition (\ref{normaliz}) implies that the weak charges are the generators of a {\it weak isospin} SU(2) group. In SU(3) space,  $\theta$ determines the orientation of the weak SU(2) group with respect to the {\it strong} SU(2) group, which is determined by the medium strong interactions which break SU(3) to the familiar isotopic spin symmetry. In the absence of the medium strong interactions, one could identify the weak isospin group with the isospin symmetry and strange particles would be stable under weak decays. 

The interplay of the weak and medium-strong interactions to determine the value of $\theta$ proved to be far reaching. It has remained in the present unified theory in the form of a misalignment between the weak isospin subgroup of the flavor symmetry and the quark mass matrix, which arises from the spontaneous symmetry breaking of the weak isospin gauge symmetry.

\section{The angle as a dynamical effect of strong vs weak interactions} 

Cabibbo entertained for sometime the idea that the value of the weak angle, $\theta$, could be determined by theoretical considerations. The fact that the angle indicates the direction of the weak isospin group in SU(3) space could be seen as a kind of {\it spontaneous magnetization} in SU(3) space and its value should arise as a solution of a self-consistency equation for the symmetry breaking parameter, presumably an SU(3) symmetric equation. This led to the problem of finding the {\it natural solutions} of equations invariant under a given group, $G$. The problem was tackled theoretically by L. Michel and L. Radicati\cite{LMLR}, who investigated the natural minima in SU(3), always finding trivial minima corresponding to $\theta=0$ or $\pi$. 
Cabibbo and myself\cite{CABMA68} extended the analysis to the chiral symmetry group  SU(3)$\times$SU(3) with two possible symmetry breaking structures, transforming as:
\be 
(3,{\bar3}) \oplus ({\bar 3},3)~{\rm or}~(8,1) \oplus (1,8)
\ee
but again {\it we found} only trivial results.

In modern terms, computing the Cabibbo angle means to determine theoretically the structure of the quark mass matrix, which, with three quark flavour{\it s}, would correspond to the first choice in the previous equation. Attempts in this direction have met with some success\cite{Fritzsch:1977za}, which amounts to justify the {\it empirically valid} relation:
\be
\sin\theta \simeq \sqrt{\frac{m_u}{m_s}} 
\ee
between $\theta$ and the up and strange quark masses, but a really convincing theory has not emerged yet and $\theta$ is still to be considered an undetermined constant of Nature. 

Historically, the attempt to compute the Cabibbo angle was one of the motivations that led to the discovery of the GIM mechanism. One should not give up the idea that sometimes we shall be able to compute the pattern of symmetry-breaking quark masses and therefore to compute the Cabibbo angle. The more so, since, after the discovery of neutrino oscillations, the problem reproposes itself for the neutrino mass matrix.

Michel and Radicati ideas have been later used to justify the natural symmetry breaking patterns of Unified and Grand Unified theories.

\section{Closing up on Cabibbo theory}  

From its very publication, the Cabibbo theory has been seen as a crucial development. It indicated the correct way to embody lepton-hadron universality and it enjoyed a heartening phenomenological success, which in turns indicated that we could be on the right track towards a fundamental theory of the weak interactions. 

The authoritative book by A. Pais\cite{Pais:1986nu}, in its chronology, quotes the Cabibbo theory among the most important developments in post-war Particle Physics.

In the {\it History of CERN}, J.Iliopoulos\cite{ilio96} writes:{ \it There are very few articles in the scientific literature in which one does not feel the need to change a single word and CabibboÕs is definitely one of them. With this work he established himself as one of the leading theorists in the domain of weak interactions.} 

\section{Post- Cabibbo developments: a unified, renormalizable, electroweak theory} 

\begin{figure}[htb]
\begin{center}
\includegraphics[scale=.65]{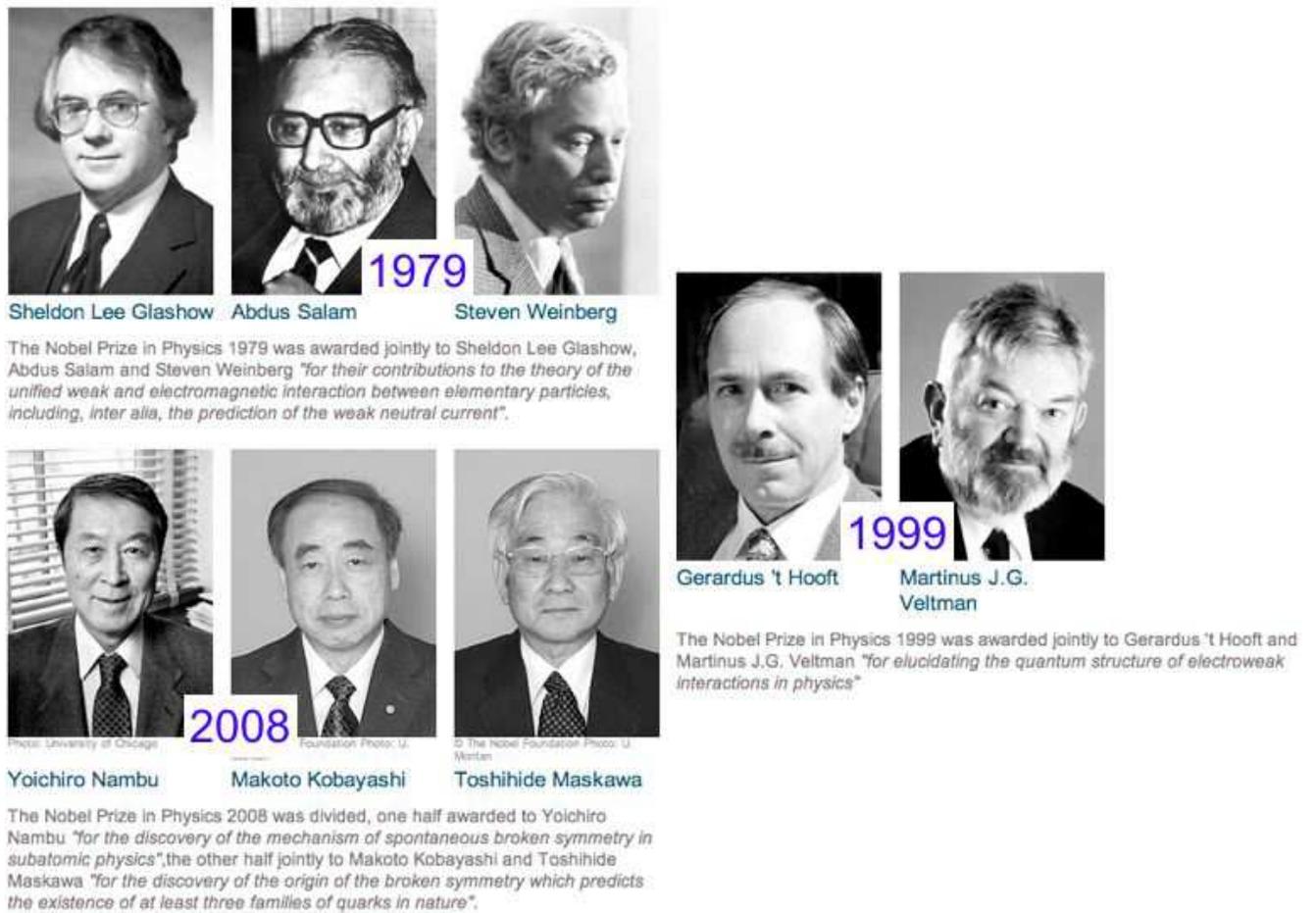}
\caption{{\small  Nobel Prize winners who contributed to the theory of the unified electroweak interactions; Cabibbo theory has been a crucial step towards this great achievement.}}
\label{nobel}
\end{center}
\end{figure}

Eight Nobel Prizes (Fig. \ref{nobel}) have been given for the theory of the unified electroweak interactions pioneerd by S. L. Glashow\cite{Glashow:1961tr}, S. Weinberg\cite{Weinberg:1967tq}  and A. Salam\cite{salam}. The Cabibbo theory has been a crucial step towards this great achievement. 

%
%%%%%%%%%%%%@@@
%%%%%%%%%@@@@
\vskip0.5cm%

Post-Cabibbo developments are summarized in the following.
\begin{itemize}
\item
 The introduction of the charmed quark by S. Glashow, J.~Iliopoulos and L.~Maiani\cite{GIM70} made it possible to extend the Weinberg-Salam theory to hadrons, restoring lepton-quark symmetry and predicting hadronic weak neutral currents without strangeness change at about the same rate as charged currents; the suppression of the strangeness changing neutral currents fixes the mass scale of charmed particles, in agreement with experimental observation;
\item
G. t'  Hooft and M. Veltman, in 1972, proved the renormalizability of the spontaneously broken (via the Higgs mechanism) gauge theory\cite{'tHooft:1972fi};
\item
Adler anomalies in $SU(2)\times U(1)$ were the last obstacle towards a renormalizable electroweak theory and they were proven  to cancel between quark (fractionally charged and in three colors) and lepton doublets, by C.~Bouchiat, J.~Iliopoulos and P.~Meyer\cite{Bouchiat:1972iq}.
\end{itemize}

\section{CP violation}

1973. Kobayashi and Maskawa discovery\cite{Kobayashi:1973fv}: three left-handed quark doublets allow for one CP violating phase in the quark mixing matrix, since known as the CKM matrix;

1976. S. Pakvasa and H. Sugawara\cite{Pakvasa:1975ti}
and L. Maiani\cite{Maiani:1975in}, show that the phase agrees with the observed CP violation in K decays and (LM) leads to  vanishing neutron electric dipole at one loop;

1986. I. Bigi and A. Sanda\cite{Bigi:1981qs}
   predict direct CP violation in B decay;

2001. Belle\cite{Abe:2001xe} and BaBar\cite{Aubert:2001sp} discover CP violating mixing effects in B-decays.

\section{New Challenges}
Problems which were on the table at the beginning of our story, %namely 
the end of {\it the} 1950's, have all been solved by an extraordinary mix of theoretical inventions and experimental results. Some of the crucial steps have been described in this paper.

The proliferation of nuclear particles and resonances, initiated with the discovery of strange particles, has found an explanation in terms of more fundamental fermion fields, quarks coming in six flavours, each with three colours. The muon has found its place in the second quark-lepton generation. The fifth and sixth quarks neatly pair with the ($\nu_\tau, \tau$) lepton doublet in a third generation, necessary to explain the CP violation initially observed with particles belonging to the first and second generations. 

We understand the structure of the weak and electromagnetic currents, their renormalisation properties and the relation between leptonic, semi-leptonic and non-leptonic weak processes. The unified gauge theory of both interactions, electromagnetic and weak, has been experimentally confirmed in crucial instances, including existence and properties of the predicted, necessary, weak intermediaries. The mathematical consistency of the theory requires, by the way, precisely the lepton-quark simmetry which is so prominent in the spectrum of the elementary fermions.

Neutrino oscillations have been observed, in particular where they are required to support our understanding of the way the Sun {\it shines}. We now know that neutrinos have masses, similarly to quarks and charged leptons, and that the phenomenon of fermion mixing, discovered by Cabibbo, is quite general, although we do not know yet how to predict its structure. 

The description of the basic strong interactions with an asymptotically free gauge theory based on the colour symmetry is, perhaps, the most unexpected and most spectacular development of the second half of the last century. It has allowed for crucial  {\it quantitative} tests of the strong interactions, in the short distance region where we can apply perturbative methods. Non-perturbative calculations based on the numerical simulation of QCD in a space-time lattice, have produced highly non trivial results in the large distance, strongly interacting, regime. One instance is the calculation of the axial couplings of the pseudo-scalar mesons, although, admittedly, we are still far from a systematic understanding of this domain. A gauge description of all fundamental interactions, including gravity, {\it strongly suggests the existence }of a unified theory encompassing all interactions, realising the dream of Albert Einstein.

With the turn of the Century, we have a new panorama of problems and challenges and a new machine, the Large Hadron Collider at CERN, to explore a new energy domain, ranging from 100 to above 1000 GeV=1 TeV. I will list only a few of the challenges which may be attacked in the new round of experiments at the LHC. This is a personal list and may well turn out to be incomplete or even irrelevant: future will tell.

The first challenge is to find the Higgs boson\cite{Higgs:1966ev}. The Higgs boson is needed for the unified electro-weak theory to agree with Nature, validating the idea that symmetry breaking particle masses arise from the spontaneous breaking of the gauge symmetry.  At the same time, this mechanism gives a vision of the quantum vacuum which may help us to explain new phenomena in the universe at large: inflation, chaotic universe, etc..

Find the supersymmetric particles. The unification of forces requires a symmetry to relate different spins: this is {\it Supersymmetry}, a fermion-boson symmetry discovered in 1974 at CERN by J. Wess and B. Zumino\cite{Wess:1974tw} and in Russia by  D. Akulov and V. Volkov\cite{Akulov:1974xz}. 

There are arguments, related to the so-called hierarchy problem of fundamental scales, that suggest the presence of the supersymmetric partners of the known particles in the TeV range\cite{Davier:1979hr}, possibly within reach of the LHC. 

Indications for a form of stable matter other than we know, protons, {\it neutrons}, electrons and neutrinos, come independently from the existence of non-luminous matter, gravitationally observed in the Universe.  In fact, the data on the primordial abundance of helium and other light nuclei limit the abundance of baryonic matter to a few percent of the total mass and neutrinos are definitely too light. The origin of the  {\it  dark matter} is {\it definitely} one of the most prominent puzzles of present physics.
A neutral, very long lived, supersymmetric partner surviving from the hot Big Bang could be a natural candidate to be the constituent of the dark matter in the Universe.

Finally, the search for extra space-dimensions. String formulations of Quantum Gravity are not consistent in 3+1 dimensions. Curved extra-dimensions are needed. How small is their radius ? Can LHC high energy particles get into and map for us the new dimensions?

\section{Cabibbo: Leading the Roma school }

Nicola settled in Roma {\it La Sapienza} in 1966, moved to Roma {\it Tor Vergata} for few years and came back to {\it La Sapienza}.
Inspired by Nicola's physical intuition, mathematical skill and personal carisma, the Rome school significantly contributed to establishing what we call today the Standard Theory of particle physics, which Nicola had greatly helped to build. A few results of these wonderful years.
 \begin{itemize}
\item
The parton-model description of $e^+ e^-$ annihilation into hadrons\cite{Cabibbo:1970as};
\item
the first calculation of the electroweak contribution to the muon anomaly\cite{Altarelli:1972nc};
\item
field theoretic description of the parton densities in hadrons\cite{Altarelli:1973ff};
\item
QCD prediction of a phase transition from hadrons into deconfined quarks and gluons starting from the limiting temperature introduced by R. Hagedorn\cite{Cabibbo:1975ig};
\item
CP and T reversal violation in the oscillations of three flavored neutrinos\cite{Cabibbo:1977nk};
\item 
upper and lower bounds to the Higgs boson and heavy fermion masses in Grand Unified theories\cite{Cabibbo:1979ay};
\item
parton analysis of the heavy quark $\beta$-decay spectrum (allowing one of  the most precise determinations of the CKM mixing parameters)\cite{Cabibbo:1978sw} \cite{Altarelli:1982kh};
\item
lattice QCD calculation of weak parameters with lattice QCD\cite{Cabibbo:1983xa};
\item 
with G. Parisi, Cabibbo proposed and realised a parallel supercomputer for lattice QCD calculations\cite{Bacilieri:1984ai}. The APE supercomputers and their subsequent evolutions have played an important role in elucidating basic QCD in the non-perturbative regime.
\end{itemize}
%A nice picture of Nicola Cabibbo in these years is reported in the following figure.

%
%%%%%%%%%%%%@@@
\begin{figure}[htb]
\begin{center}
\includegraphics[scale=.55]{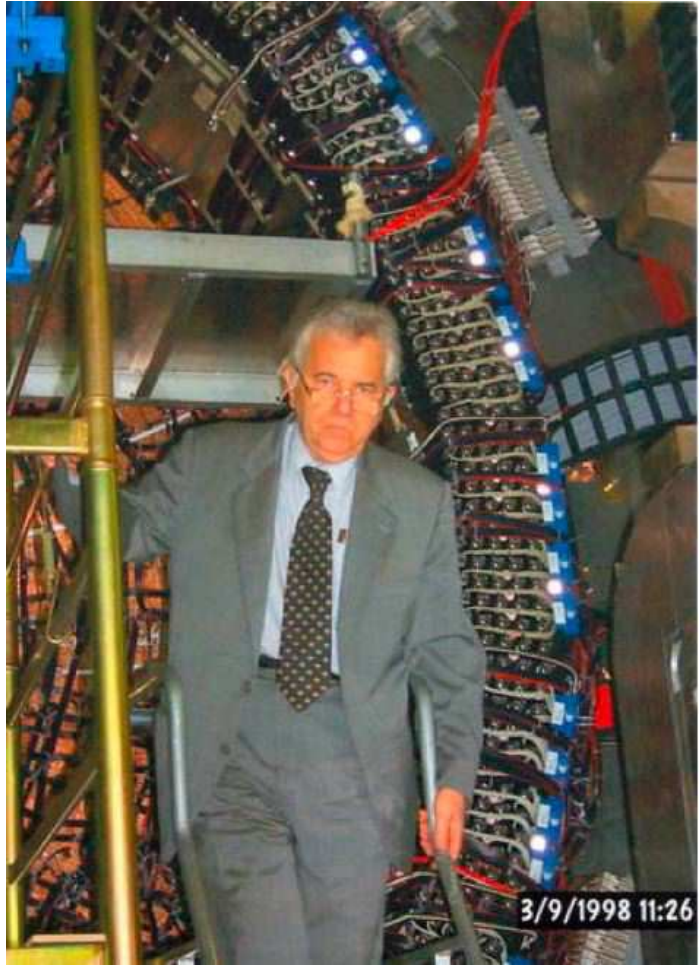}
\caption{{\small  Nicola Cabibbo in 1998, visiting the KLOE detector in Frascati. Courtesy of Andrea Cabibbo.}}
\end{center}
\label{nobelwinners}
\end{figure}
%%%%%%%%%@@@@

\section{Nicola Cabibbo: Science Manager, teacher and friend}

Nicola played an overall important role in the Italian scientific life of the turn of the century, as:

Member of Academia Nazionale dei Lincei and of the American Academy of Science;

President of Istituto Nazionale di Fisica Nucleare: 1983-1992;

President of Ente Nazionale Energie Alternative: 1993 -1998;

President of the Pontifical Academy of Science: from 1993;

He held these important positions with vision, managerial skill and universally appreciated integrity. 

Nicola liked to teach and he continued to do so until his very last months. 
Like all great minds, he could find simple arguments to explain the most difficult concepts. 
His students were fascinated by his simplicity, gentle modes and sense of humour. 
So we did, all of us we who had the privilege to be his collaborators and friends.

\end{document}